\documentclass[doublecol]{epl2}
\usepackage{amsmath}
\usepackage{amssymb}
\usepackage{mathrsfs}
\usepackage[mathscr]{euscript}

\begin{document}
\title{A non-BCS mechanism for superconductivity in underdoped cuprates via
attraction between spin vortices}

\author{P. A. Marchetti\inst{1}, F. Ye\inst{2},
Z. B. Su\inst{3}, L. Yu\inst{4,3}}

\institute{

  \inst{1} Dipartimento di Fisica ``G. Galilei'', INFN, I-35131 Padova,
  Italy

  \inst{2} College of Material Science and Optoelectronics Technology,
  Graduate University of Chinese Academy of Science, Beijing 100049,
  China

  \inst{3} Institute of Theoretical Physics, Chinese Academy of
  Sciences, Beijing 100190, China

  \inst{4} Institute of Physics, Chinese Academy of Sciences, Beijing
  100190, China }

\date{\today }

\abstract{ We propose a non-BCS mechanism for superconductivity in
  hole-underdoped cuprates based on a gauge approach to the {\it t-J}
  model. The gluing force is an attraction between spin vortices
  centered on the empty sites of two opposite N\'eel sublattices,
  leading to pairing of charge carriers. In the presence of these pairs,
  a gauge force coming from the single occupancy constraint induces, in
  turn, the pairing of the spin carriers. The combination of the charge
  and spin pairs gives rise to a finite density of incoherent hole
  pairs, leading to a finite Nernst signal as precursor to
  superconductivity. The true superconducting transition occurs at an
  even lower temperature, via a 3D XY-type transition. The main features
  of this non-BCS description of superconductivity are consistent with
  the experimental results in underdoped cuprates, especially the
  contour plot of the Nernst signal.  }
% \pacs{ 71.10.Hf, 11.15.-q, 71.27.+a}

\maketitle

In this Letter we propose a new mechanism of superconductivity in
hole-underdoped High T$_c$ cuprates, using the spin--charge gauge
approach to the 2D {\it t-J} model, describing the CuO
planes \cite{jcmp}. In this approach the disturbance of hole doping
on the antiferromagnetic (AF) background is systematically
considered, giving rise to spin vortices dressing the charge
excitation (fermionic spinless holon).  At the same time,
due to these vortices the spin excitation (bosonic spin 1/2 spinon)
acquires a finite gap, leading to a short range (SR) AF order. The
interplay of that SR order with the dissipative motion of charge
carrier results in a metal-insulator crossover, a pronounced
phenomenon in the underdoped cuprates. A number of peculiar
features of cuprates in the normal state can be well explained
within this scheme \cite{jcmp}. Here this approach is generalized
to consider the superconducting state.

The gluing force of the superconducting mechanism is an attraction
between spin vortices on two opposite N\'eel sublattices, centered
around the empty sites (holes), and we propose a three-step scenario: At
the highest crossover temperature, denoted as $T_{ph}$, a finite density
of incoherent holon pairs are formed.  We propose to identify this
temperature with the experimentally observed (upper) pseudogap (PG)
temperature, where the in-plane resistivity deviates from the linear
behavior.  However, the holon pairing alone is not enough for
superconductivity to appear.  Due to the no-double occupation
constraint, there is a gauge interaction between holon and spinon,
through which the spin vortex attraction induces in turn the formation
of spin-singlet (RVB) spinon pairs with a reduction of the spinon
gap. At the intermediate crossover temperature, denoted as $T_{ps}$, a
finite density of incoherent spinon RVB pairs are formed, which,
combined with the holon pairs, gives rise to a gas of incoherent
preformed hole pairs. We propose to identify this temperature with the
experimental crossover corresponding to the appearance of the
diamagnetic and Nernst signal. Finally, at an even lower temperature,
the superconducting transition temperature $T_c$, both holon pairs and
RVB pairs, hence also the hole pairs, become coherent.  The proposed
superconducting mechanism is not of the BCS-type, and it involves a gain
in kinetic energy (for spinons) coming from the $J$-term of the spin
interactions. The main features of this non-BCS description of
superconductivity are consistent with the experimental results in
underdoped cuprates, especially the contour plot of the diamagnetic and
Nernst signal \cite{ong1,ong2}.

{\it The spin-charge gauge approach} \cite{jcmp} relies on the following
key ideas: {\it 1)} We decompose the hole operator of the {\it t-J}
model as $c_\alpha = h^*z_\alpha$, where $h$ is a spinless fermionic
holon, carrying charge, while $z_\alpha$ is a spin 1/2 bosonic spinon,
carrying spin, together with an emergent slave-particle gauge field
($A_\mu$), minimally coupled to holon and spinon, taking care of the
redundant $U(1)$ degrees of freedom coming from the spin-charge
decomposition. With this choice the no-double occupation constraint is
automatically satisfied via the Pauli principle. {\it 2)} In 2D (and 1D)
one can add a ``statistical'' spin flux ($e^{i \Phi^s}$) to $z$ and a
``statistical'' charge flux ($e^{i \Phi^h}$) to $h$, provided they
``compensate'' each other in an appropriate way so that the product
$e^{-i \Phi^h} h^* e^{i \Phi^s} z$ is still a fermion. The introduction
of these fluxes in the lagrangian formalism is materialized via the
Chern-Simons gauge fields. We find the optimal charge and spin
statistical fluxes in the mean-field approximation (MFA)
\cite{jcmp}. The effect of the optimal spin flux is to attach a
spin-vortex to the holon, with opposite chirality on the two N\'eel
sublattices, the rigidity holding up a vortex being provided by the AF
background.  These vortices take into account the long-range distortion
of the AF background caused by the insertion of a dopant hole, as first
discussed in \cite{ss}; they are naturally associated with the semionic
nature of the spin flux (see.\cite{jcmp}) and are reminiscent of
Laughlin's vortices of Ref. \citeonline{la}.  Neglecting
$A$-fluctuations, the leading terms of the Hamiltonian can be written
as:
\begin{eqnarray}
  \label{tJ} H&=& \sum_{\langle ij \rangle} (-t) \text{AM}_{ij}
  h^{*}_i h_j e^{i
    (\Phi^h_j-\Phi^h_i)} + h.c. \nonumber\\
  &&+J(1-h^*_ih_i-h^*_jh_j) (1-|\text{AM}_{ij}|^2)\nonumber\\
&&  + J h^*_ih^*_jh_jh_i |\text{RVB}_{ij}|^2,
\end{eqnarray}
where $\text{AM}_{ij}= z^\dagger_i e^{-i \Phi^s_i}e^{i\Phi^s_j} z_j$ is
a kind of Affleck-Marston spinon parameter \cite{am} and
$\text{RVB}_{ij} =\sum\epsilon_{\alpha \beta} z_{i \alpha} z_{j \beta}$
is an RVB spinon singlet order parameter.  {\it 3)} We use the following
improved MFA: in the first term of (\ref{tJ}) we make the MFA
$\langle\text{AM}_{ij}\rangle \approx 1$, while in the second term we
replace the hole density by its average, and in the normal state we
neglect the third term because of being higher order in doping
($\delta$). Notice that such MFA correctly reproduces the critical
exponents of the 1D {\it t-J} model \cite{np}, when dimensionally
reduced (the spin-vortices becomes kink strings in 1D). A
long-wavelength treatment of the second term in (\ref{tJ}) leads to a
CP$^1$ spinon nonlinear $\sigma-$model with an additional term coming
from the spin flux,
\begin{equation}
\label{mass} \tilde J (\nabla \Phi^s)^2 z^\dagger z,
\end{equation}
where $\tilde J =J(1-2\delta)$ and $\partial_\mu \Phi^s
(x)=\epsilon_{\mu \nu}\partial_\nu \sum_{j} (-1)^{|j|} \Delta^{-1} (x -
j) h^*_jh_j $ with $ \Delta$ the 2D lattice laplacian. {\it 4)} In a
quenched treatment of spin vortices, self-consistently justified if
their density is not too small, we derive the MF expectation value
$\langle(\nabla\Phi^s)^2\rangle = m_0^2 \approx 0.5\delta |\log
\delta|$, which opens a mass gap for the spinon, consistent with AF
correlation length at small $\delta$ extracted from the neutron
experiments \cite{ke}. This gap is also crucial for eliminating an
overcounting of low-energy degrees of freedom often encountered in
slave-particle approaches, giving rise to problems in the computation of
thermodynamic quantities\cite{hl}. In fact, because of the spinon gap,
the low-$T$ thermodynamics in this approach is essentially dominated by
the gapless holons, while the contribution of transverse and scalar
gauge fluctuations to the free energy almost canceling each
other\cite{ma}.  {\it 5)} In the parameter region corresponding to the
PG ``phase'' of the cuprates \cite{nota} the optimal charge flux is
$\pi$ per plaquette and via Hofstadter mechanism it converts the
spinless holons $h$ into Dirac fermions with a small Fermi Surface (FS)
$\epsilon_F \sim t\delta$. Their dispersion is defined in the Magnetic
Brillouin Zone (MBZ), which we choose as a union of two square regions,
denoted as R(ight) and L(eft) centered at $\vec{Q}^{R}=(\pi/2,\pi/2)$
and $\vec{Q}^{L}=(-\pi/2,\pi/2)$, respectively; these momenta are also
the centers of the holon FS.  Increasing doping or temperature one
reaches the crossover line $T^* \approx t/(8 \pi) |\log \delta|$
entering the ``strange metal phase'' (SM) of the cuprates \cite{nota}
where the optimal charge flux per plaquette is $0$ instead of $\pi$ and
we recover a ``large'' FS for the charge excitations with $\epsilon_F
\sim t (1-\delta)$. {\it 6)} Holons and spinons are coupled by the
dressed gauge field $A$ giving rise to metal-insulator crossover and
overdamped resonances for holes and magnons with strongly $T$-dependent
life-time\cite{jcmp}.

Let us now turn to superconductivity (SC). The SC order parameter is
assumed to be RVB-like: $\Delta^c = \langle \sum_{\alpha \beta}
\epsilon_{\alpha\beta} c_{\alpha i} c_{\beta j} \rangle$. In our MF
treatment, neglecting gauge fluctuations it is given by the product of
$\langle \sum_{\alpha\beta}\epsilon_{\alpha\beta} z_{\alpha i} z_{\beta
  j} \rangle$ and $\langle h^*_i h^*_j \rangle$. Hence both expectation
values should be non--vanishing to have SC in MFA.

{\it Holon pairing.} In our approach the only quartic term in (\ref{tJ})
is the RVB-like term which is repulsive. However, spin vortices centered
on holons have opposite vorticity on the two N\'eel sublattices and that
produces a long-range attraction, previously neglected in the MFA; this
will be our key attractive force.  Physically it is due to the
distortion of the AF background caused by the holes. This effect in the
simplest form was first realized by S. Trugman\cite{tr} in the early
days of High Tc studies, who correctly pointed out that putting two
holes next to each other on two N\'eel sublattices would save energy
2$J$.  We include this effect in MFA by introducing a term coming from
the average of $z^\dagger z$ in (\ref{mass}):
\begin{equation}
\label{zh}
\tilde{J}  \langle z^* z \rangle \sum_{i,j} (-1)^{|i|+|j|} \Delta^{-1}
(i - j) h^*_ih_i h^*_jh_j,
\end{equation}
where $\Delta$ is the 2D lattice laplacian.  In the static approximation
for holons (\ref{zh}) describes a 2D Coulomb gas with coupling constant
$g=\tilde J \langle z^\dagger z \rangle$, where $\langle z^\dagger z
\rangle \sim (\Lambda^2+m_0^2)^{1/2}-m_0 $ with $\Lambda \approx 1$ as a
UV cutoff, and charges $\pm 1$ depending on the N\'eel sublattice. For
2D Coulomb gases with the above parameters a pairing appears for a
temperature $T_{ph} \approx {g} /2\pi $, which turns out to be inside
the SM ``phase'' \cite{nota0}. Hence the whole PG ``phase'' lies below
$T_{ph}$. However, we will discuss only the SC arising from the PG
phase, anticipating that extrapolation to SM phase will introduce only
quantitative changes.

To implement the above ideas it is useful to distinguish holons on the
two N\'eel sublattices, denoting them as $a$ and $b$, and the two square
regions R and L of the MBZ.  The corresponding holons are called $a^{L},
b^{L}$ and $a^R,b^R$, respectively. We also measure the momentum from
the left and right centers of the two regions $\vec{Q}^{R},\vec{Q}^{L}$.
Since not all vortices form pairs, a finite screening effect persists
and the gas of vortices still have a finite correlation length, which we
denote as $\xi \approx (J k_F)^{-1/2}$ \cite{dl}. We keep track of the
screening effect by replacing (in the long wavelength limit) $
\Delta^{-1}$ in (\ref{zh}) by an effective potential between vortices on
different N\'eel sublattices, whose Fourier transform extends around 0
in a region of order $ k_F$ and there it is given by $V_{eff}(q) \approx
g/(q^2 + \xi^{-2})$. The coupling between the L and R regions occurs
only through $V_{eff}$, but since $V_{eff}(\vec{Q}^{R}+\vec{Q}^{L})\ll
V_{eff}(\vec 0)$ we neglect this coupling.

We define:
$$
\Delta^h_\alpha (\vec{p})= \int d^2 q V_{eff}(\vec{q}-\vec{p})
\left\langle b^{\alpha}_{-\vec{q}} a^{\alpha}_{\vec{q}}\right\rangle
$$
with $\alpha=R,L$. As in Ref.\cite{sus1,sus2}, the $d$-wave pairing
symmetry is composed of two $p$-wave pairing within the left and
right Dirac cones \cite{nota1} in the form

\begin{eqnarray}
\label{eq:45}
\Delta^h_\alpha (\vec{k}) = \left\{
  \begin{array}{ll}
    \Delta^h(k) (k_x-k_y), \text{ if }\alpha=R, \\
    \Delta^h(k) (-k_x-k_y), \text{ if }\alpha=L.
  \end{array}
\right.
\end{eqnarray}
We adopt the BCS approximation and the energy spectrum takes the
following four-branch form
\begin{eqnarray}
  \label{hsp}
  E_\alpha (\vec{k})  = \pm \sqrt{(\mu\pm|2 t{\vec{k}}|)^2+|\Delta^h_\alpha (\vec{k})|^2},\quad \alpha=R,L.
\end{eqnarray}
Neglecting branches not crossing FS and following the procedure
developed for a spin-wave attraction mechanism in Ref.\cite{sus1,sus2},
we obtain for the solution of the gap equation on the FS

\begin{eqnarray}
\label{Deltah} \Delta^h \equiv \Delta^h(k_F) \approx  g \xi
\exp\{-{ c\over g \xi^2 k_F }\},
\end{eqnarray}
with a constant $c$. As common for non-weakly coupled attractive Fermi
systems, the MF temperature at which $\Delta^h$ becomes non-vanishing
should be identified with the pairing temperature $T_{ph}$.  One
reinserts gauge fluctuations and recovers gauge invariance in the nodal
approximation by a standard recipe as discussed in \cite{ft}: One
introduces explicitly the coordinate-dependent argument of $\Delta^h$,
$\arg \Delta^h$. The first component of the nodal holon is multiplied by
$e^{-i {1 \over 2} \arg \Delta^h}$, while the second component by $e^{i
  {1 \over 2}\arg \Delta^h}$. The resulting field is a slave-particle
gauge-invariant, hence physical, ``nodon''.  One then reinserts
appropriately the gauge-invariant vector field $a_\mu = A_\mu - {1\over
  2}\partial_\mu \arg \Delta^h$ by the Peierls substitution.  It is now
easy to derive the low-energy effective action in the nodal
approximation for $a$, as a variant of QED${}_3$.

{\it spinon RVB pairing}. This can be materialized by the RVB term
in (\ref{tJ}) combined with the gauge attraction. The RVB term is
repulsive and in MFA of holons it is irrelevant in absence of the
holon pairing, but it becomes relevant as soon as  holon pairing
appears. In fact, as shown below, the gauge attraction then favors
an RVB condensation of short-range spinon pairs. Introducing a complex
RVB-Hubbard-Stratonovich gauge field $\Delta^s_{\langle
ij\rangle}$ and treating the holon pair in MFA, the RVB-term in
(\ref{tJ}) becomes:
\begin{equation}
\label{sg}
 \sum_{\langle ij\rangle} \frac{\Delta^s_{\langle ij\rangle} \Delta^{s*}_{\langle ij\rangle}}{2 J |\langle h_ih_j\rangle|^2}
  + \Delta^{s*}_{\langle ij\rangle}\epsilon_{\alpha\beta} z_{\alpha i} z_{\beta j } +h.c.
\end{equation}

In the continuum limit following \cite{ft} we present
$\Delta^s_{\langle ij\rangle}$ as a product of a
space-independent, but direction dependent modulus factor times a
space-dependent phase.

Neglecting gauge and phase fluctuations and assuming rotational
invariance, from (\ref{sg}) we derive the modified four-branches spinon dispersion:
\begin{equation}
\label{sd}
{\omega(\vec{k}) = \pm \sqrt{(m_s^2 -
|\Delta^s|^2) + (\tilde{J}|\vec k| \pm
|\Delta^s|)^2,}}
\end{equation}
where $m_s=\tilde{J} m_0$. The positive branches of the dispersion
(\ref{sd}) are similar to those found in a plasma of relativistic
fermions \cite{wel}. It suggests the following interpretation: if
$|\Delta^s|\neq 0 $ the spinon system contains a gas of RVB spinon
pairs, an analogue of Coulomb neutral pairs in the relativistic
plasma, either in the plasma phase, if $\langle \Delta^s\rangle$=0
or in a condensate, if $\langle \Delta^s\rangle \neq 0$.  For a
finite density of spinon pairs there are two (positive energy)
excitations, with different energies, but the same spin and
momenta. They are given, {\it e.g.}, by creating a spinon up and
by destructing  a spinon down in one of the  RVB pairs. Notice
that the minimum at $\tilde{J} |\vec k|= |\Delta^s|$ in the lower
branch is like the roton minimum in superfluid helium and has an
energy lower than $m_s$; it implies a backflow of the gas of
spinon-pairs dressing the ``bare spinon''. Hence RVB condensation
would lower the spinon kinetic energy. However, to make it occur
one needs the gauge contribution.

For $\langle \Delta^s\rangle \neq 0$ the global slave particle
symmetry is broken from $U(1)$ to ${\bf Z}_2$, due to the
condensation of the charge 2 $e$ holon and RVB pairs. The
Anderson-Higgs mechanism then implies a gap $\sim |\Delta^s|$ for
the gauge field $A_\mu$.  To exhibit it explicitly one calculates
from (\ref{sg}) the gauge effective action, obtained by
integrating out the spinon. The result up to quartic terms for the
lagrangian density is
\begin{eqnarray}
&\;L_{eff}(a,\partial \chi)=\frac{1}{2 \sqrt{(m_s^2-|\Delta^s|^2)}}
[(\partial_\mu a_\nu-\partial_\nu a_\mu)^2+\\&\; {|\Delta^s|^2}
(2(a_0-\partial_0 \chi)^2+(\vec a-\vec\nabla \chi)^2)] \nonumber
\label{leff}
\end{eqnarray}
with $\chi={1 \over 2} (\arg \Delta^h-\arg \Delta^s)$ a phase field,
slave-particle gauge-invariant, hence physical, whose gradient
describes the potential of standard magnetic vortices. From
previous results one can derive the gap equation for $|\Delta^s|$
in the continuum MFA, hence neglecting $\chi$, at $T=0$,
(setting
$\Lambda=1$),
\begin{eqnarray}
\label{gapeq}
&&
\frac{\tilde J}{m_s^2}-\frac{1}{J|\langle
h_ih_j\rangle|^2} \nonumber\\
&=& \int d \omega d \vec{k}
 \frac{4 \tilde J^2 k^2}{(\omega^2+ m_s^2 + \tilde J^2 k^2)^2-4 \tilde J^2
  k^2 |\Delta^s|^2. }
\end{eqnarray}
In the l.h.s. of Eq.~\eqref{gapeq} the first term originates from the
gauge action due to the lowering of the spinon mass ($m_s \rightarrow
(m_s^2-|\Delta^s|^2)^{1/2}$), while the second term comes from the
original repulsive Heisenberg term. The r.h.s.  is due to spinons; the
contribution of the spectrum of gauge bosons is neglected as
subdominant.  Whereas in the slave-boson approach the RVB pairs are made
of fermions and the Heisenberg term is attractive, so the pair-formation
is BCS-like, in our approach the RVB pairs are made of bosons, and in
our chosen representation the Heisenberg term is repulsive, while the
pair formation arises from the decrease in the free energy of spinons,
via the lowering of their mass gap, induced by holon-pairing through the
gauge field.  Notice that the leading part of the original Heisenberg
term is used to provide the AF action for the spinons, using the
identity (holding for bosonic spinons) $|AM|^2+|RVB|^2=1$, see
eq. (\ref{tJ}).  Only the subleading term proportional to the holon-pair
density is used to obtain the formation of a finite density of RVB-pairs
in (\ref{gapeq}), so the derived superconductivity appears a little bit
in the spirit of Laughlin's Gossamer superconductivity \cite{lau}.  One
easily realizes that in (\ref{gapeq}) a non-vanishing solution for
$|\Delta^s|$ is possible only if the second term is sufficiently small,
{\it i.e.}, for sufficiently large MF holon pair density. Extension to
finite $T$ is straightforward including also the contribution from the
gauge bosons (with an approximate form of the effective action due to
holons) and roughly estimating $|\langle h_ih_j\rangle| \sim \Delta^h
k_F J^{-1}$ up to rescaling.  The temperature at which $|\Delta^s|$
becomes non-vanishing is $T_{ps}$, not yet the true condensation
temperature, $T_c$.  Notice that from (\ref{gapeq}) it follows
necessarily $T_{ps}<\;T_{ph}$.  The $\delta$ behavior obtained from
(\ref{Deltah}) and the finite $T$ version of (\ref{gapeq}) for the
crossover temperatures $T_{ph}$ and $T_{ps}$, as well as the contour
plot for different $\Delta^s$ are shown in Fig. 1.
\begin{figure}[htbp]
\centerline{\includegraphics[width=4.5cm]{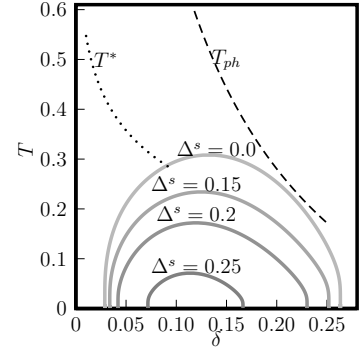}}
\caption[]{The $T-\delta$ phase diagram of the mean field gap equation
  of spinon for different values of MF spinon pairing $\Delta^s$ (gray
  lines) which could be compared with different levels of the Nernst
  signal; $\Delta^s=0$ is $T_{ps}$.  (The curves at high dopings are not
  quantitatively reliable as they do not take into account the crossover
  to SM).  The dashed line is $T_{ph}$, the ``upper PG crossover
  temperature''.  The dotted line is the crossover temperature between
  the PG and SM phases, $T^*$ (see also \cite{nota0}). The temperature
  and $\Delta^s$ are in units of $J$.}
\end{figure}

{\it SC transition}. The SC transition appears as a XY-type transition
for magnetic vortices. In fact in the gauged XY model (\ref{leff}) if
the coefficient, $|\Delta^s|^2$ , of the Anderson-Higgs mass term for
$a$ is sufficiently small the angular field $\chi$ fluctuates so
strongly that it does not produce a mass gap for $a_\mu$ and $\langle
e^{i \chi}\rangle=0$.  This is the Coulomb phase of the gauged XY (or
Stueckelberg) model, where a plasma of magnetic vortices-antivortices
appears. In the presence of a temperature gradient a perpendicular
magnetic field would induce an unbalance between vortices and
antivortices, giving rise to a Nernst signal.  Therefore we conjecture
that this phase of model(\ref{leff}) corresponds to the region in the
phase diagram characterized by a non-SC Nernst signal and a comparison
of the experimental phase diagram in \cite{ong1,ong2} and the one
derived in our model (see Fig. 1) supports this idea.

For a sufficiently large coefficient $|\Delta^s|^2$, on the other hand,
we are in the broken symmetry phase; the fluctuations of $\chi$ are
exponentially suppressed and $\langle e^{i \chi}\rangle \neq 0$ at $T=0$
or there is quasi-condensation at $T>0$ and the gauge field is gapped.
One can prove that, due to the fluctuations of the field $\arg
\Delta^h$, in our approach a gapless gauge field is inconsistent with
the coherence of holon pairs in PG, {\it i.e.,} coherent holon pairs
{\it cannot coexist} with incoherent spinon pairs. On the other hand,
due to the QED-like structure of holons-gauge action, the gauge field
cannot be gapped by condensation of holon pairs alone; only the
condensation of RVB spinon pairs at the same time can provide a gap to
gauge fluctuations. Thus as soon as $e^{i \chi}$ (quasi-)condenses the
same occurs to $\langle h_ih_j\rangle$ so that SC emerges, since the SC
order parameter is $\Delta^c=\Delta^s/\langle h_ih_j\rangle$. It follows
that $T_c<\;T_{ps}$ and the transition is of XY-type being triggered by
the behavior of the XY field $\chi$.

A few comments on comparison of the present proposal with other models
on superconductivity in cuprates are in order. It is clear that our
proposal differs in an essential way from the traditional BCS-Eliashberg
approach\cite{sch}, no matter whether the electron-phonon interaction or
the antiferromagnetic fluctuations serve as the pairing glue. The SC
transition occurs here in a similar way as in the preformed pairs
formalism\cite{ek}. From the physical point of view, this approach is an
implementation of the basic idea advocated by P.W. Anderson, attributing
superconductivity to the strong correlation effects in doped Mott
insulators\cite{and1,bas,and2}. It shares some similarities with other
formalisms exploring the same physical idea, with, however, some
substantial differences. Both in the standard slave-boson \cite{leeRMP}
and in the bosonic-RVB phase-string\cite{weng1,weng2} approaches the
Nernst effect and SC occur due to Bose-Einstein condensation (BEC) of
bosonic holons. Since BEC persists for arbitrary small density in these
approaches both Nernst effect and SC at $T=0$ occur as soon as the
long-range AFO disappears. The same also happens in the standard
``preformed pair'' approaches \cite{ek}, due to the persistence of
condensation of pairs in the extreme BEC limit. Instead in our approach
the repulsive interaction between spinons prevents the appearance of the
Nernst effect below a critical doping, and the hole pairing occurs only
when the holon pair density is sufficiently large to ``force'' the RVB
spinon pairing via gauge coupling, while an even higher doping at $T=0$
is necessary to get SC. Similar ``critical'' dopings also appear in the
phase-fluctuation approach of \cite{tes}, the main physical difference
with ours being in that approach nodes appear in the Nernst phase,
whereas in ours a finite FS still persists and nodes appear only in the
SC phase. Furthermore, in previous approaches no clear evidence of our
additional crossover $T^*$ appears, as distinct from our $T_{ph}$ (often
denoted there as $T^*$).

{\it Remark}: Our approach presents 3 distinct crossover lines
which different authors have alternatively considered as the
``pseudogap'' crossover: the highest one in $T$, $T_{ph}$ where
holons start to pair reducing the spectral weight of the hole
\cite{mg}; a second lower one, $T_{ps}$ where incoherent hole
pairs are formed, mainly affecting the magnetic properties since a
finite FS still persists; and a third one, $T^*$, intersecting
$T_{ps}$, where one crosses from a large to a small holon FS, with
a complete suppression of the spectral weight for the holes in the
antinodal region, with physical effects observable experimentally
both in transport and thermodynamics \cite{ma}. The first two
crossovers have mainly a magnetic origin, in the formalism
described by the spin flux $\Phi^s$, while the third one has a
charge origin, due to the charge flux $\Phi^h$, and appears only
in two-dimensional bipartite lattices, see \cite{jcmp}.

{\it Conclusions:} We have proposed a non-perturbative pairing mechanism
in High Tc cuprates. We believe it captures the most pronounced
characteristics of these compounds: a strong interplay of
antiferromagnetism and superconductivity. The same Heisenberg
interaction derived from the strong on-cite Coulomb repulsion is
responsible for both antiferromagnetism and SC pairing: the leading term
of that interaction gives rise to antiferromagnetism, while its
sub-leading term providing the paring glue due to vortex-antivortex
attraction on the AF background. Compared with other proposals on
pairing mechanisms, it describes this interplay in a more natural way.
Although many details of our approach are admittedly conjectural, the
mechanism of SC proposed here is rather complete in its main structure
and has the following appealing features: 1) It is not of simple BCS
structure, in agreement with some experimental data in underdoped
cuprates \cite{deutscher,van,wen}. 2) SC appears only at a finite doping
above the critical point where long range AF disappears. 3) It allows
vortices in the normal state, as in the preformed pair scenario,
supporting a Nernst signal. 4) The appearance of two positive branches
in the spinon dispersion relation for a suitable spinon-antispinon
attraction induces a similar structure for the magnon dispersion around
the AF wave vector \cite{mg}, reminiscent of the hour-glass shape of
spectrum found in neutron experiments\cite{hour}. Furthermore, since the
spinon gap has a maximum in $\delta$, the energy of the magnon resonance
is also expected to have a maximum. 5) In the SC state the gauge gap
destroys the Reizer singularity which is responsible for the anomalous
$T$-dependent life-time of the magnon and electron resonances in the
normal state. Therefore one expects that in the SC state, these
resonances become sharper at the superconducting transition. The
compositeness of excitations within the gauge approach, the holon-spinon
and spinon-antispinon composites with a gauge glue coming from the
single-occupancy constraint, proved to be essential in interpreting the
transport\cite{jcmp} and thermodynamical\cite{ma} properties of cuprate
superconductors, turns out to be also crucial for the superconductivity
to actually occur. Furthermore, the nature of the three-step
crossovers/transition needed for superconductivity, proposed in this
approach: BCS-like for holon-pair formation, kinetic-energy driven for
RVB spinon-pair formation and XY-like for the true superconducting
transition, can be seen as specific experimental predictions of this
approach. A more complete presentation of our approach to
superconductivity will appear in \cite{mysy}.


\begin{thebibliography}{10}
\bibitem{nota} In the $T-\delta$ phase diagram we identify the crossover
  from PG to SM with the inflection point in resistivity and the broad
  peak in the specific heat coefficient $\gamma$.
\bibitem{nota0} The expression found for $T_{ph}$ (and $T^*$) cannot be
  extrapolated to very small doping, because of the failure of the
  quenched approximation for spin-vortices producing the spinon gap.
\end{thebibliography}
\end{document}